\documentclass[aps,twocolumn,preprintnumbers,amsmath,amssymb,nofootinbib,superscriptaddress,notitlepage]{revtex4-1}

\usepackage{graphicx,color}
\usepackage{relsize}
\usepackage{slashed}
\usepackage[normalem]{ulem}
\usepackage{tabu}
\usepackage{rotating}
\usepackage{bigstrut}
\usepackage{makecell}
\usepackage[dvipsnames]{xcolor}

\usepackage{amsmath}
\usepackage{amssymb,bm}
\usepackage{amsthm}
\usepackage{mathrsfs}
\usepackage{array}
\usepackage[all]{xy}
\usepackage{eufrak}
\usepackage{euscript}
\usepackage{enumerate}
\usepackage{slashed}
\usepackage{hyperref}
\usepackage{subfigure} 
\usepackage{mathtools}

\usepackage{soul}
\graphicspath{{figs/}}

\hypersetup{pdftex,colorlinks=true,linkcolor=blue,citecolor=blue,menucolor=black,urlcolor=blue,filecolor=blue}

\newcommand{\swu}{\affiliation{School of Physical Science and Technology,
	Southwest University, Chongqing 400715, China}}

\newcommand{\q}{\mathbf{q}}
\newcommand{\intq}{\int_{|\mathbf{q}|<q_{\max }}}

\newcommand{\maojun}[1]{{#1}}

\begin{document}

\title{Reevaluating the $a_1(1420)$ enhancement and its molecular partners in the low-lying axial-vector meson spectrum}

\author{Mao-Jun Yan}
\email{yanmj0789@swu.edu.cn}
\swu


\author{Chun-Sheng An}
\swu

\author{Cheng-Rong Deng}
\swu

\date{\today}


\begin{abstract} 
We investigate possible axial-vector poles associated with the
near-threshold $K^{\ast}\bar K$ interaction, driven by the
Weinberg-Tomozawa term and unitarized through the Bethe-Salpeter
equation. In the isovector sectors with
$I^G(J^{PC})=1^\pm(1^{+\mp})$, we find second-sheet poles below the
$K^{\ast}\bar K$ threshold and examine their possible manifestations
in the $a_1(1420)$ structure and its possible partner $b_1(1400)$.
The corresponding $f_0(980)\pi$ and $\phi\pi$ invariant-mass
distributions are compared with the COMPASS and BESIII data,
respectively. Within the present framework, the enhancements around 1.4 GeV may be qualitatively understood through the interplay between the triangle-singularity mechanism and the pole-enhanced $K^\ast\bar K$ final-state interaction, although a quantitative description requires a more complete treatment of coupled channels and production dynamics.
\end{abstract}

\maketitle

\section{Introduction}

Thanks to the substantial progress in experimental particle physics
over the past two decades, an increasing number of new hadronic states
have been observed since 2003. Many of them exhibit properties that
are difficult to accommodate within the conventional quark model of
quark-antiquark mesons and three-quark baryons, and have therefore
been considered as candidates for exotic hadrons; see, e.g.,
Refs.~\cite{Chen:2016qju,Hosaka:2016pey,Esposito:2016noz,
Lebed:2016hpi,Ali:2017jda,Olsen:2017bmm,Guo:2017jvc,
Albuquerque:2018jkn,Liu:2019zoy,Guo:2019twa,Brambilla:2019esw,
Chen:2022asf}.

Compared with the heavy-hadron sector, where clear exotic candidates
such as the $Z_c(3900)^\pm$ have been observed
\cite{BESIII:2013ris,Belle:2013yex}, the situation in the light-quark
sector remains less conclusive. A recent example is the
$\eta_1(1855)$ structure observed by BESIII in
$J/\psi\to\gamma\eta\eta^\prime$
\cite{BESIII:2022riz,BESIII:2022zel}. Its favored quantum numbers,
$J^{PC}=1^{-+}$, are exotic in the conventional $q\bar q$ quark
model. Although various interpretations of the $\eta_1(1855)$ have
been proposed, its underlying structure remains under discussion
\cite{Lacock:1996vy,Dudek:2010wm,Qiu:2022ktc,Chen:2022qpd,
Shastry:2022mhk,Yan:2023vbh,Yang:2022rck,Dong:2022cuw}.

Another interesting example is the $a_1(1420)$ structure observed by
the COMPASS Collaboration in the diffractive reaction
$\pi N\to\pi\pi\pi N$ \cite{COMPASS:2018uzl}. A peak was found in
the $f_0(980)\pi$ channel at
$1414^{+15}_{-13}\,\rm{MeV}$, with a width
$\Gamma\simeq153^{+8}_{-23}\,\rm{MeV}$
\cite{COMPASS:2018uzl}. Although initially interpreted as a possible
new meson, it was subsequently pointed out that the structure can be
generated by a triangle singularity (TS) associated with the
$K^{\ast}\bar K K$ loop
\cite{Zhao:2013,Mikhasenko:2015oxp,Aceti:2016yeb,
COMPASS:2020yhb}. In this mechanism, the $a_1(1260)$ couples to
$K^{\ast}\bar K$, followed by $K^{\ast}\to K\pi$, while the
remaining $K\bar K$ pair couples to the $f_0(980)$.

More recently, an $a_1(1420)$-like contribution has also been
reported in the partial-wave analysis of
$\tau^-\to\pi^-\pi^-\pi^+\nu_\tau$ by Belle
\cite{Rabusov:2023tna}. The extracted mass and width,
$1387.8\pm0.3\,\rm{MeV}$ and $109.2\pm0.6\,\rm{MeV}$,
respectively, are somewhat smaller than the COMPASS values.
Differences between the two analyses may be related, among other
effects, to the different production environments and coherent
background contributions. These observations provide additional
motivation for investigating the dynamics underlying the
$a_1(1420)$ structure.

The role of the TS mechanism in the $a_1(1420)$ signal has been
extensively studied, from the original theoretical proposals
\cite{Mikhasenko:2015oxp,Aceti:2016yeb}
\footnote{The possible connection of the $a_1(1420)$ to a triangle
singularity was first pointed out by Qiang Zhao at the HADRON 2013
Conference in Nara, where the COMPASS result was first reported
\cite{Guo:2019twa}.}
to phenomenological analyses incorporating three-body dynamics over
a broad energy range
\cite{COMPASS:2020yhb,Sakthivasan:2024uwd,Sakthivasan:2026bph}.
The TS is intrinsically tied to a specific kinematic configuration
and produces its strongest effect in a relatively narrow energy
region close to the $K^{\ast}\bar K$ threshold. This makes the
near-threshold region particularly suitable for investigating
whether nonperturbative $K^{\ast}\bar K$ dynamics can further affect
the observed line shape.

A TS and a nearby pole can both generate pronounced
structures in an invariant-mass distribution, although their analytic
origins are different. A TS corresponds to a logarithmic branch-point
singularity \cite{Guo:2019twa},
$A_{\rm TS}\sim\log(s-s_{\rm TS})$, whereas a resonance or
near-threshold state is associated with a pole of the scattering
amplitude. Consequently, the observation of a peak alone is generally
insufficient to distinguish between kinematic and dynamical
mechanisms, and their possible interplay should be examined at the
amplitude level.

Motivated by this observation, we investigate the near-threshold
$K^{\ast}\bar K$ interaction within the chiral unitary approach.
The pseudoscalar-vector ($PV$) interaction is described at leading
order in the chiral expansion by the Weinberg--Tomozawa (WT) term
and unitarized through the Bethe--Salpeter equation. We find a
second-sheet pole below the $K^{\ast}\bar K$ threshold in the
isovector negative-$G$-parity sector and investigate its possible
manifestation in the $a_1(1420)$ region.
\footnote{A meson-meson molecular component may in general mix with
compact multiquark configurations
\cite{Yan:2021wdl,Deng:2022cld,Deng:2021gnb,Peng:2023lfw}.
In particular, a $K^{\ast}\bar K$ component with $J^{PC}=1^{++}$
may couple to tetraquark configurations with the same quantum
numbers \cite{Chen:2015fwa,Sundu:2017xct,Huang:2026cig}.}
Our analysis focuses on the near-threshold region. Additional
coupled channels, such as $\rho\pi$ and $f_0(980)\pi$, as well as
more complete three-body dynamics, may become important in a global
description over a broader energy range and are beyond the scope of
the present calculation.

To examine the observable consequences, we calculate the
$f_0(980)\pi$ invariant-mass distribution by combining the
nonperturbative $K^{\ast}\bar K$ final-state interaction with the
triangle transition $K^{\ast}\bar K\to f_0(980)\pi$, and compare the
resulting line shape with the COMPASS data. Rather than performing a
global multi-parameter fit, we concentrate on the near-threshold
region and investigate how the $K^{\ast}\bar K$ rescattering modifies
the TS-induced line shape.

We further extend the analysis to the isovector positive-$G$-parity
sector, $I^G(J^{PC})=1^+(1^{+-})$, which contains the established
$b_1(1235)$ meson \cite{ParticleDataGroup:2024cfk}. Since the
$K^{\ast}\bar K$ interaction is governed by the same WT kernel in
this sector, we find a partner second-sheet pole below the
$K^{\ast}\bar K$ threshold, which we denote as $b_1(1400)$.
\footnote{The notation $b_1(1400)$ is related to the state denoted
as $\tilde b_1(1420)$ in Ref.~\cite{Guo:2017jvc}.}
Interestingly, an enhancement around $1400\,\rm{MeV}$ is present
in the $\phi\pi^0$ invariant-mass distribution measured by BESIII
in $J/\psi\to\eta\phi\pi^0$
\cite{BESIII:2018ozj,BESIII:2023zwx}. A TS mechanism has also been
discussed for this process. We therefore investigate whether the
observed structure can be described through the interplay between
the TS mechanism and the pole-enhanced $K^{\ast}\bar K$
final-state interaction.

This article is organized as follows. In
Sec.~\ref{sec:II}, we introduce the chiral unitary framework and the
WT interaction for $S$-wave $PV$ scattering. In
Sec.~\ref{sec:Isovector}, we discuss the isovector
$K^{\ast}\bar K$ interaction and its manifestations in the
$f_0(980)\pi$ and $\phi\pi^0$ invariant-mass distributions measured
by COMPASS and BESIII, respectively. We then briefly discuss
extensions to other flavor sectors before summarizing our results.

\section{pseudoscalar-vector mesons scattering}\label{sec:II}

The $S$-wave interaction between $K^{\ast}$ and $\bar{K}$ given by the WT term reads~\cite{Roca:2005nm}
\begin{eqnarray}
V(s)=\frac{\epsilon \cdot \epsilon^{\prime}}{8 F_\pi^2} C\left[3
s-2(M^2+m^2)-\frac{1}{s}(M^2-m^2)^2\right],
\label{Eq:wt}
\end{eqnarray}
where
$F_{\pi}$ corresponds to the pion decay constant ($\simeq 92.1$~MeV), and
$\epsilon(\epsilon^{\prime})$ is the four-vector polarization of the incoming (outgoing) $K^{\ast}$ meson. Additionally, we denote $M$,
and $m$ as the masses for the $K^{\ast}$
and $K$ mesons, respectively. The coefficient $C=-1$ in isovector scattering has been given
in Ref.~\cite{Roca:2005nm}. 
Note that the transition of $K^{\ast}\bar{K}$ is attractive in the isovector sector, which is similar to those of the $N \bar{K}$ and $K_1 \bar{K}$ interactions, studied in Refs.~\cite{Lu:2022hwm,Yan:2023vbh}, which resulted in states attributed to the structures $\maojun{\Sigma(1430)}$ and $\pi_1(1600)$, respectively.

The $K^{\ast}\bar K$ scattering amplitude is unitarized
through the Bethe-Salpeter equation in its on-shell
factorized form~\cite{Oller:1997ti}
\begin{eqnarray}
T=(1-V \hat{G})^{-1} V\, ,
\label{Eq:BS}
\end{eqnarray}
where $V$ represents the kernel matrix, whose elements are determined by Eq.~\eqref{Eq:wt} above. Meanwhile, $\hat{G}=G_{\ell} (1+\frac{q^2_l}{3 M^2_l})$ is the diagonal matrix with elements $G_{\ell}$ corresponding to the loop function for the $K^{\ast}\bar{K}$ channel. On the one hand, given that $G_{\ell}$ is ultraviolet (UV) divergent, it can be regularized by introducing a cutoff in the three-momentum $q_{\max}$, such that
\begin{widetext}
\begin{eqnarray}
    G_{\ell} &=& i \int \frac{d^4 q}{(2 \pi)^4} \frac{1}{q^2-m_{\ell}^2} \frac{1}{(q-P)^2-M_{\ell}^2}
    = \intq \frac{d^3 q}{(2 \pi)^3} \frac{\omega_1(\q)+\omega_2(\q)}{2 \omega_1(\q) \omega_2(\q)} \frac{1}{\left(P^0\right)^2-\left(\omega_1(\q)+\omega_2(\q)\right)^2}\, ,
    \label{Eq:Gl_cut}
\end{eqnarray}
\end{widetext}
where $\omega_1(\q)=\sqrt{M_l^2+|\q|^2}$ and $\omega_2(\q)=\sqrt{m_l^2+|\q|^2}$. Here, $m_{\ell}$ and $M_{\ell}$ represent the masses of the pseudoscalar and vector mesons, respectively, involved in the loop, and $P$ the total four-momentum of those mesons $\left(P^2=s\right)$. On the other hand, we can also adopt the dimensional regularization scheme to deal with the UV divergence. In this case, the loop-function assumes the following form~\cite{Oller:1998zr,Oller:2000fj}
\begin{widetext}
\begin{align}
G_{\ell}^{\mathrm{DR}}(s)= &\, \frac{1}{16 \pi^2}\left[\alpha_{\ell}(\mu)+\log \frac{M_{\ell}^2}{\mu^2}+\frac{m_{\ell}^2-M_{\ell}^2+s}{2 s} \log \frac{m_{\ell}^2}{M_{\ell}^2}\right. \nonumber\\
&\left.+\frac{p_{\ell}}{\sqrt{s}}\left(\log \frac{s-m_{\ell}^2+M_{\ell}^2+2 p_{\ell} \sqrt{s}}{-s+m_{\ell}^2-M_{\ell}^2+2 p_{\ell} \sqrt{s}}
+\log \frac{s+m_{\ell}^2-M_{\ell}^2+2 p_{\ell} \sqrt{s}}{-s-m_{\ell}^2+M_{\ell}^2+2 p_{\ell} \sqrt{s}}\right)\right], \label{eq:Gdr} \,
\end{align}
\end{widetext}
where $\alpha_{\ell}(\mu)$ is a subtraction constant that absorbs any changes in the scale of the regularization, denoted by $\mu$, and $p_{\ell}$ corresponds to the three-momentum of the mesons in the center-of-mass (c.m.) frame, given by 
\begin{eqnarray}
 p_{\ell}=\frac{1}{2 \sqrt{s}}{\sqrt{\left(s-\left(M_l+m_l\right)^2\right)\left(s-\left(M_l-m_l\right)^2\right)}}\,.  
 \label{Eq:pcm}
\end{eqnarray}
In this work, we set $\mu=1000\,\rm{MeV}$. Furthermore, the subtraction constant $\alpha_{\ell}(\mu)$ for each $PV$ loop is obtained by matching Eqs.~\eqref{Eq:Gl_cut} and \eqref{eq:Gdr} at the corresponding $PV$ threshold. 
\maojun{For the single-channel $K^\ast\bar K$ calculation considered here, matching ~\eqref{Eq:Gl_cut} and \eqref{eq:Gdr} at the $K^\ast\bar K$ threshold gives $\alpha(\mu)=-1.28$ for $q_{\max}=800~\mathrm{MeV}$ and $\alpha(\mu)=-1.63$ for $q_{\max}=1000~\mathrm{MeV}$, with $\mu=1000~\mathrm{MeV}$.}

In order to search for poles in the $T$-matrix, we look at its solutions on the energy complex plane. As a result, we should address the interplay of the different Riemann sheets stemming from the branch cut in the $G_{\ell}$ loop function. For instance, there are two Riemann sheets for a single-channel system. In this case, the poles that show up below threshold on the real energy axis on the first (physical) Riemann sheet correspond to bound states. Poles below threshold on the second (unphysical) Riemann sheet are
conventionally associated with virtual states. In the present case,
where the pole may lie away from the real energy axis, we simply refer
to it as a second-sheet pole below the $K^{\ast}\bar K$ threshold. Thus, denoting $G^{\rm I}_{\ell}(s)$ and $G^{\rm II}_{\ell}(s)$ as the loop-functions in the first and second Riemann sheets, respectively, we have
\begin{align}
	G_{\ell}^{\rm II}(s)= G^{\rm I}_{\ell}(s)+i \frac{ p_{\ell}}{4 \pi \sqrt{s}}.
\end{align}

\section{Isovector spectrum}\label{sec:Isovector}

\subsection{ $a_1(1420)$}\label{sec:IsovectorA}

In the isospin symmetric basis,  the Riemann sheets in this system are labeled as RS$_{\pm}$, where the subindex denotes the sign of Im$q_1$, with $q_1$ the c.m. momenta of $K^{\ast}\bar{K}$.
As a result, we find poles located at $\sqrt{s_p} = 1336 - i\,130 \,\rm{MeV}$ using $\maojun{\alpha(\mu)=-1.28}$, and at $\sqrt{s_p} =  1331 - i\,77$~MeV for $\maojun{\alpha(\mu)=-1.63}$ on RS$_{-}$, where \maojun{the finite $K^{\ast}$ width is neglected}. 
It is worth emphasizing that, in this strict single-channel limit without including the ($K^{\ast}$) width, there is no physical phase space for decay. Therefore, the sizable imaginary part of the pole should not be interpreted as a conventional decay width. Rather, it arises from the analytic continuation of the energy-dependent WT interaction onto the second Riemann sheet. The imaginary part thus characterizes the location of the pole in the complex-energy plane.

\begin{figure}[htb]
	\centering
\includegraphics[scale=0.450]{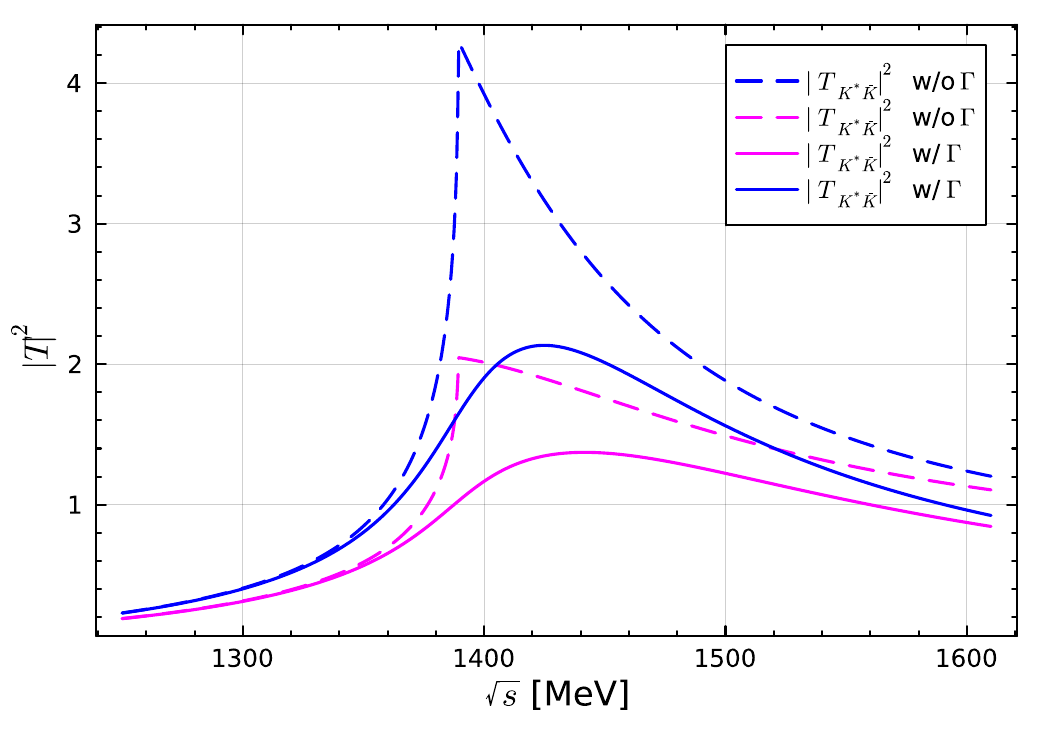}
	\caption{Modulus squared of $T$-matrix elements. The dashed and solid lines correspond to the solutions of Eq.~\eqref{Eq:BS}, considering the vector meson width as zero and finite, respectively. The \maojun{blue} lines are determined by choosing a subtraction constant for the loops involved, matching Eqs.~\eqref{Eq:Gl_cut} and \eqref{eq:Gdr} at threshold with  $\maojun{\alpha(\mu)=-1.63}$, while the magenta lines are determined with $\maojun{\alpha(\mu)=-1.28}$. }\label{AmpSq1}
\end{figure}

These two poles yield non-trivial structures around $\sqrt{s}=1400\,\rm{MeV}$ in the modulus squared of the $T$-matrix, as shown in Fig. \ref{AmpSq1}. On the other hand, when the finite width of the $K^{\ast}$ is
incorporated through a Breit--Wigner prescription, the right-hand
cut is smeared and the pole can no longer be determined in the same
way. The cusp structure is then smoothed into a broad peak close to
the mass of the $a_1(1420)$ structure reported by COMPASS in
Ref.~\cite{COMPASS:2018uzl}. Notice that this dynamically generated pole is the flavor partner of $W_{c1}$ \footnote{\maojun{The notation $W_{c1}$ is used in Refs.~\cite{Peng:2023lfw,Zhang:2024fxy,Ji:2025hjw, Nakamura:2026xax,Germani:2026yhe,Dong:2026qbr,Shi:2026hwn,Wang:2026wai}  to denote a predicted near-threshold C-parity partner of the $Z_c(3900)$, associated with the $D\bar D^{\ast}$ interaction.}} in the hidden charm sector \cite{Peng:2023lfw,Zhang:2024fxy,Ji:2025hjw, Nakamura:2026xax,Germani:2026yhe,Dong:2026qbr,Shi:2026hwn,Wang:2026wai}.

In both cases, the $K^{\ast}\bar K$ amplitude exhibits a
pronounced structure in the mass region relevant to the
$a_1(1420)$. This suggests that the $a_1(1420)$ structure reported by COMPASS
may reflect an interplay between the TS mechanism and the
pole-enhanced $K^{\ast}\bar K$ final-state interaction.

\begin{figure}[htb]
	\centering
\includegraphics[scale=0.350]{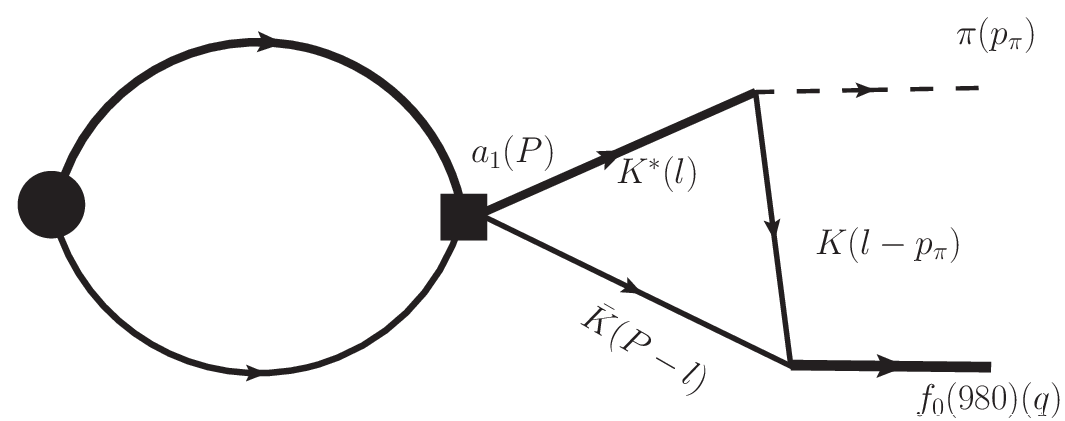}
	\caption{The $K^{\ast}\bar{K}$ transition to $f_0(980)\pi$ via a triangle diagram. The black dot and rectangle represent the $K^{\ast}\bar{K}$ production rate and the unitarized scattering amplitude, respectively. The bold, solid, and dashed lines represent the vector mesons, kaon, and pion, respectively. }\label{a1_WT}
\end{figure}

\begin{figure}[tb]
	\centering
\includegraphics[scale=0.450]{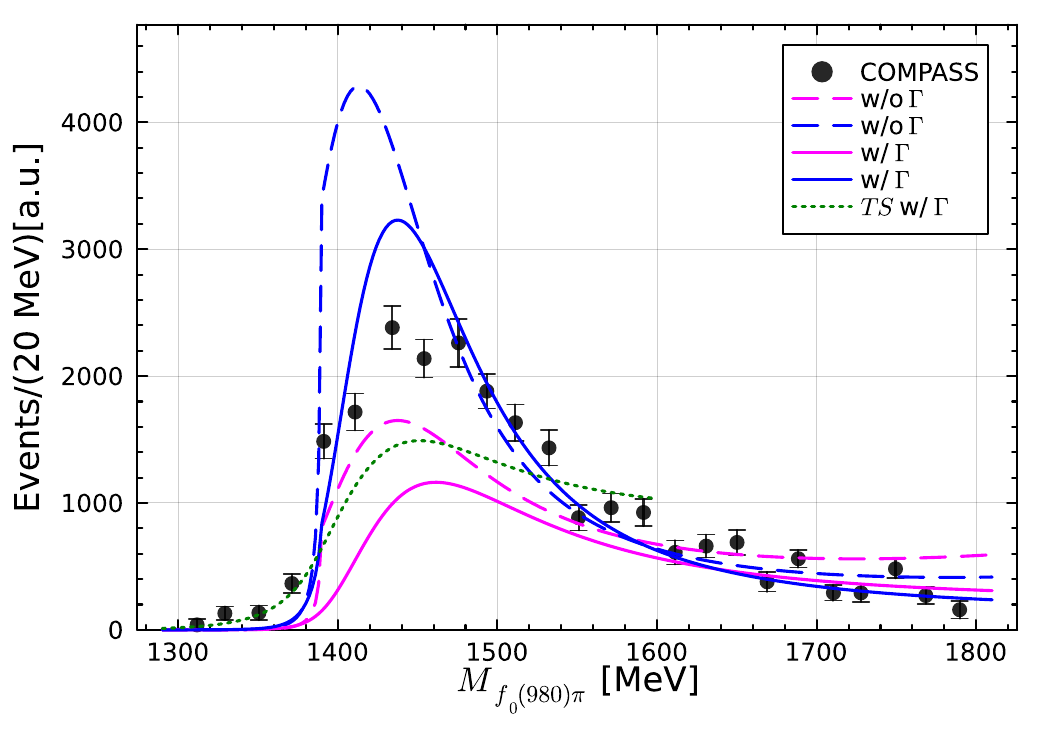}
	\caption{Numerical results for Eq.~\eqref{Eq:xsec_f0} corresponding to the $f_0(980)\pi$ invariant mass distribution compared to the data from COMPASS \cite{COMPASS:2020yhb}. The magenta and blue lines correspond to the cases of $\maojun{\alpha(\mu)=-1.28}$ and  $\maojun{\alpha(\mu)=-1.63}$ in the G-loop function, while the solid and dashed lines stand for including the width of $K^{\ast}$ or not, respectively, where the width of $K^{\ast}$ is imposed into the loop integral in the form of Breit-Wigner. \maojun{ The green dotted line represents the TS-only contribution
with the finite $K^{\ast}$ width and the spectral function of $f_0(980)$.} 
 }\label{Amp_f0pi}
\end{figure}

We can test this hypothesis by calculating the differential cross section $d N_{a1}/dM_{f_0 \pi}$ as a function of the invariant mass of the $f_0(980)\pi$ pair and comparing it with the corresponding COMPASS data. Considering the Feynman diagram illustrated in Fig.~\ref{a1_WT}, the effective Lagrangian in the transitions of $K^{\ast}$ decaying into $K\pi$ and $K\bar{K}$ coupling to $f_{0}(980)$ are 
\begin{eqnarray}
    \mathcal{L}_{K^{\ast}K\pi}&=& g_1 K^{\ast,\mu}\left(K\partial_{\mu}\pi-\partial_{\mu}K\pi\right),\\
    \mathcal{L}_{f_0 K\bar{K}}&=& g_2 f_0 K \bar{K},
\end{eqnarray}
with coupling constants $g_1=\frac{m_{\rho}}{2F_{\pi}}$ \cite{Yan:2026srb}  and $g_2\maojun{=3.63\,\rm{GeV}}$ quoted from Ref. \cite{Oller:1998zr}, where $m_{\rho}$ is $\rho$ meson mass. Consequently, the corresponding amplitude can be written as
\begin{equation}
\mathcal{M}_{a1}=P_{a_1}\,\hat{G}_{K^*\bar{K}} T_{K^*\bar{K}}\mathcal{M}^{TS}_{a_1}\, .
\label{Eq:amp}
\end{equation}
Here, $\hat G_{K^{\ast}\bar K}T_{K^{\ast}\bar K} $ accounts for the nonperturbative $K^{\ast}\bar K$ rescattering prior to the transition into the $f_0(980)\pi$ final state, while \(\mathcal M^{TS}_{a_1}\) describes the subsequent triangle transition \maojun{i.e., by omitting the $K^\ast\bar K $
rescattering factor $\hat G_{K^\ast\bar K}T_{K^\ast\bar K}$}. The two factors therefore represent distinct dynamical stages of the production process and should not be regarded as independent additive contributions.
\maojun{For the TS-only curve in Fig.~\ref{Amp_f0pi}, we use $\mathcal M_{a_1}^{\mathrm{TS-only}}=P_{a_1}\mathcal M_{a_1}^{\mathrm{TS}}$, i.e., the $K^\ast\bar K$ rescattering factor $\widehat G_{K^\ast\bar K}T_{K^\ast\bar K}$ is omitted rather than set to unity. Its overall normalization is chosen independently for illustrative comparison with the data.}
\begin{widetext}
\begin{eqnarray}
    \mathcal{M}^{TS}_{a_1}&=&\int \frac{d^4 l}{\left( 2\pi\right)^4}\frac{g_1 g_2\epsilon_i \delta^{ij}\left( 2\vec{p}_{\pi}-\vec{l}\right)_{\maojun{j}}}{\left( l^2-m_{K^{\ast}}^2\right) \left[ \left(P-l\right)^2-m_{\bar{K}}^2 \right] \left[\left( l-p_{\pi}\right)^2-m_{K}^2 \right]},\label{eq:TSa1}
\end{eqnarray}
\end{widetext}
\maojun{Here, $\epsilon_i$ denotes the polarization of the $a_1$ meson, while $P_{a_1}$ parameterizes the short-distance production of the $K^{\ast}\bar K$ pair. The latter is assumed to vary slowly over the energy region considered and is therefore treated as a constant that can be absorbed into the overall normalization.

Near the $K^\ast\bar K$ threshold, the nearly on-shell intermediate $K^\ast\bar K$ pair has a small relative three-momentum. In the $a_1$ rest frame, the spatial momentum $\vec{l} $ of the intermediate $K^\ast$ is therefore small compared with the outgoing pion momentum $\vec{ p}_\pi$. We thus neglect $\vec{l}$ in the numerator of the triangle integral in Eq.~(\ref{eq:TSa1}), retaining the leading near-threshold contribution while keeping the full loop-momentum dependence in the propagator denominators. This approximation is adopted to describe the triangle-singularity enhancement in the near-threshold region.}

Therefore, the differential cross section $d \tilde{N}_{a_1}/dM_{\rm inv}$ is given by
\begin{equation}
\frac{d \tilde{N}_{a_1}}{dM_{ f_0 \pi}} = \frac{|\vec{q}_\pi|}{8\pi M_{ f_0 \pi}}\,|\mathcal{M}_{a_1}|^2\, ,
\label{Eq:xsec}
\end{equation}
with $\mathcal{M}_{a_1}$  given by Eq.~\eqref{Eq:amp}, where $|\vec{q}_{\pi}|$ and $M_{ f_0 \pi}$ correspond to the magnitude of the outgoing pion momentum and the total energy in the $f_0(980)\pi$ c.m. frame, respectively.

\maojun{
To account for the finite mass distribution of the
$f_0(980)$, we employ a relativistic Breit--Wigner
spectral function,
\begin{equation}
\rho_{f_0}(s_{\pi\pi}) =
\frac{1}{\pi}
\frac{m_{f_0}\Gamma_{f_0}}
{(s_{\pi\pi}-m_{f_0}^2)^2
+m_{f_0}^2\Gamma_{f_0}^2},
\label{Eq:f0_spectral}
\end{equation}
where $s_{\pi\pi}=m_{\pi\pi}^2$,
$m_{f_0}=980$ MeV, and $\Gamma_{f_0}=50$ MeV.
The invariant-mass distribution in
Eq.~\eqref{Eq:xsec} is then convoluted with
the spectral function,
\begin{eqnarray}  
\frac{dN_{a_1}}{dM_{f_0\pi}}
&=&
\int_{4m_\pi^2}^{(M_{f_0\pi}-m_\pi)^2}
ds_{\pi\pi}\,
\rho_{f_0}(s_{\pi\pi})
\frac{|\vec q_\pi|}
{8\pi M_{f_0\pi}} \nonumber \\
&&\times\left|
\mathcal M_{a_1}
(M_{f_0\pi},s_{\pi\pi})
\right|^2,
\label{Eq:xsec_f0}
\end{eqnarray}
where
\begin{equation}
|\vec q_\pi|
=
\frac{
\lambda^{1/2}
(M_{f_0\pi}^2,s_{\pi\pi},m_\pi^2)
}{
2M_{f_0\pi}
},
\end{equation}
with $\lambda(x,y,z)=x^2+y^2+z^2-2xy-2xz-2yz$
being the K\"all\'en function.
The amplitude $\mathcal M_{a_1}$ is evaluated
at each $s_{\pi\pi}$ before performing the
integration.}

\maojun{It is worth emphasizing that, within the present framework, the $f_0(980)$ spectral function enters the production-amplitude convolution in Eq.~(\ref{Eq:xsec_f0}), whereas the pole positions are determined by the $K^\ast\bar K$ scattering amplitude in Eq.~(\ref{Eq:BS}). Thus, incorporating the $f_0(980)$ spectral function modifies the predicted production line shape but does not directly shift the pole positions within this model.}

\maojun{
In Fig.~\ref{Amp_f0pi}, we show the line shapes obtained
from Eq.~\eqref{Eq:xsec_f0} and compare them with the
COMPASS data. The magenta and blue curves correspond to
the subtraction constants $\alpha$ matched to be -1.28 and -1.63, respectively, in the
$K^{\ast}\bar K$ loop function. The solid and dashed
curves represent the results with and without the finite
$K^{\ast}$ width, respectively. In all cases, the finite
mass distribution of the $f_0(980)$ is incorporated
through the spectral function in
Eq.~\eqref{Eq:f0_spectral}.

The green dotted curve represents the TS-only
contribution, including both the finite $K^{\ast}$
width and the $f_0(980)$ spectral function.
After integration over the $f_0(980)$ mass distribution,
the TS mechanism alone produces a relatively broad
enhancement around $1400$ MeV, reflecting the
contributions from different $f_0(980)$ masses
and the corresponding triangle kinematics.
The $K^{\ast}\bar K$ rescattering introduces additional
near-threshold dynamics associated with the nearby
pole and substantially modifies the resulting line
shape. In particular, the rescattering contributions
can generate a more pronounced enhancement in the
$a_1(1420)$ region. The finite $K^{\ast}$ width
further smears the sharp structures obtained in the
stable-$K^{\ast}$ approximation.

For illustrative comparison with the COMPASS data, the overall normalization of the full-amplitude curves is fixed using the data around \(1700~\mathrm{MeV}\) as a reference. The corresponding values of \(P_{a_1}\) are \(1.33\times10^3~\mathrm{MeV}^{-1/2}\) and \(0.94\times10^3~\mathrm{MeV}^{-1/2}\) for the blue and magenta dashed curves, respectively, and \(1.70\times10^3~\mathrm{MeV}^{-1/2}\) and \(1.20\times10^3~\mathrm{MeV}^{-1/2}\) for the blue and magenta solid curves, respectively. For the green dotted TS-only curve, we use \(P_{a_1}=3.60\times10^3~\mathrm{MeV}^{-1/2}\), with the normalization chosen to match the experimental data near the \(K^\ast\bar K\) threshold. These factors affect only the overall normalization and are not obtained from a statistical fit.

The comparison suggests that the observed enhancement
may receive contributions from both the TS mechanism
and the near-threshold $K^{\ast}\bar K$ dynamics.
Although their relative importance remains model
dependent, the results demonstrate that the
$K^{\ast}\bar K$ final-state interaction can
significantly affect the observable line shape.
}

In the present calculation, coupled channels below the
$K^{\ast}\bar K$ threshold are not explicitly included
in the $T$ matrix. The finite $K^{\ast}$ width accounts
for the instability of the vector constituent and
smears the corresponding threshold structures.
Therefore, the present single-channel treatment is
not intended to provide a precision fit to the
COMPASS spectrum. The variation of the subtraction
constant and the treatment of the finite $K^{\ast}$
width illustrate the model dependence of the
predicted line shapes.

The near-threshold $K^{\ast}\bar K$ dynamics may also
influence the spectrum below the threshold,
as suggested by the behavior of the calculated
distributions in the lower-mass region. This mechanism
is distinct from the conventional $a_1(1260)$
contribution, which is commonly associated with a
$^3P_1$ $q\bar q$ state in the quark model
\cite{Cheng:2011pb} and with strong $\rho\pi$
dynamics \cite{Roca:2005nm}. The different dominant
decay channels, $\rho\pi$ for the $a_1(1260)$ and
$f_0(980)\pi$ for the $a_1(1420)$ structure
\cite{Rabusov:2023tna}, provide complementary
information on their underlying dynamics.

\subsection{ $b_1(1400)$}

In this subsection, we investigate the existence of a possible pole in the solution of Eq.~\eqref{Eq:BS}, including the channel $K^*\bar{K}$ that couples to the quantum numbers $I^G\,(J^{PC}) = 1^+\,(1^{+-})$.
Note that the WT term for the elastic $K^*\bar{K}$ scattering is identical to that in $J^{PC}=1^{++}$, in Section~\ref{sec:IsovectorA}. 
Therefore,
we can also expect the existence of a pole near $K^{\ast}\bar{K}$ threshold. Indeed, solving Eq.~\eqref{Eq:BS} with the coefficients for the WT terms in the single channel scattering, we find second-sheet poles at $\sqrt{s_p} = 1331 - i\,77$~MeV for $\maojun{\alpha(\mu)=-1.63}$, and $\sqrt{s_p} =1336 - i\,130$~MeV if we use $\maojun{\alpha(\mu)=-1.28}$. Similarly to the previous case, the corresponding amplitude exhibits
a nontrivial structure around $1400\,\rm{MeV}$. 

\begin{figure}[ht]
	\centering
\includegraphics[scale=0.350]{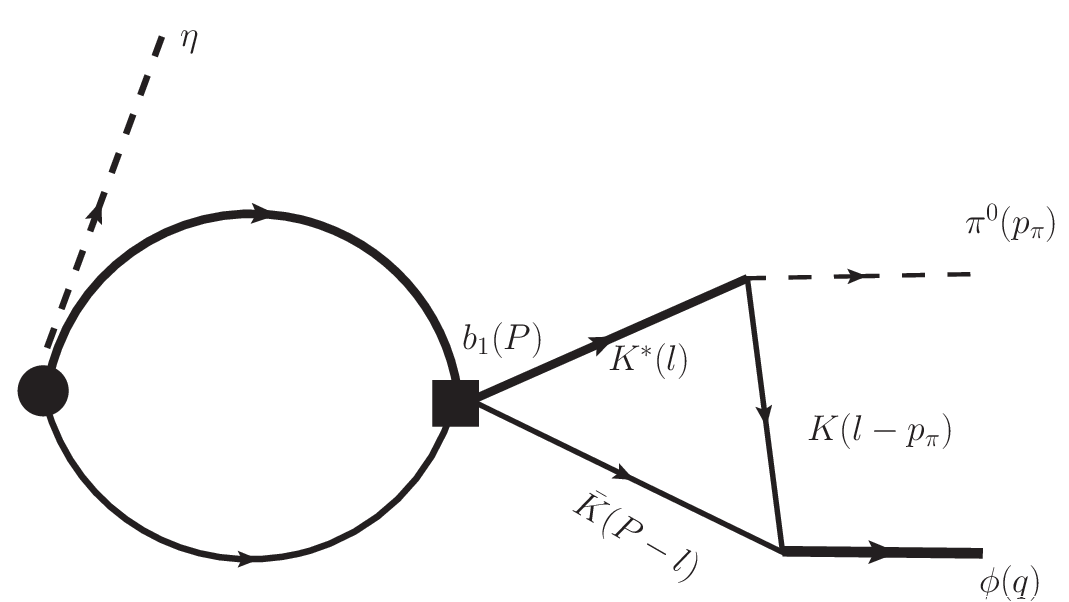}
	\caption{$K^{\ast}\bar{K}$ transition to $\phi\pi$ via the triangle diagram. The black dot and rectangle represent the $K^{\ast}\bar{K}$ production rate and the unitarized scattering amplitude, respectively. The bold, solid and dashed lines stand for the vector mesons, kaon, and pion ($\eta$), respectively.}\label{b1_WT}
\end{figure}

Since this second-sheet pole couples strongly to $K^{\ast}\bar K$, it is
natural to search for its manifestation in processes involving the
$K^{\ast}\bar K$ dynamics. A particularly interesting case is
$J/\psi\to\eta\phi\pi$, measured by the BESIII Collaboration
\cite{BESIII:2023zwx}. In Ref.~\cite{BESIII:2023zwx}, the
$M_{\pi\eta}$, $M_{\phi\eta}$, and $M_{\phi\pi}$ invariant-mass
distributions were reported, with an enhancement around
$M_{\phi\pi}\simeq 1400\,\rm{MeV}$ in the latter distribution,
which was attributed to the ``non-$\phi$ background'' component.

A TS mechanism has been proposed to account for this
structure \cite{Jing:2019cbw}. The manifestation of such a kinematic
singularity in the physical line shape, however, can be affected by
final-state interactions and by the constraints associated with
Schmid's theorem, as discussed in Refs.~\cite{Jing:2019cbw,Xiao:2024ohf}.
In view of the strong $K^{\ast}\bar K$ interaction found above, it is
therefore interesting to examine whether the observed enhancement can
be described by the interplay between the TS mechanism and the
pole-enhanced $K^{\ast}\bar K$ final-state interaction.

\begin{figure}[ht]
	\centering
\includegraphics[scale=0.450]{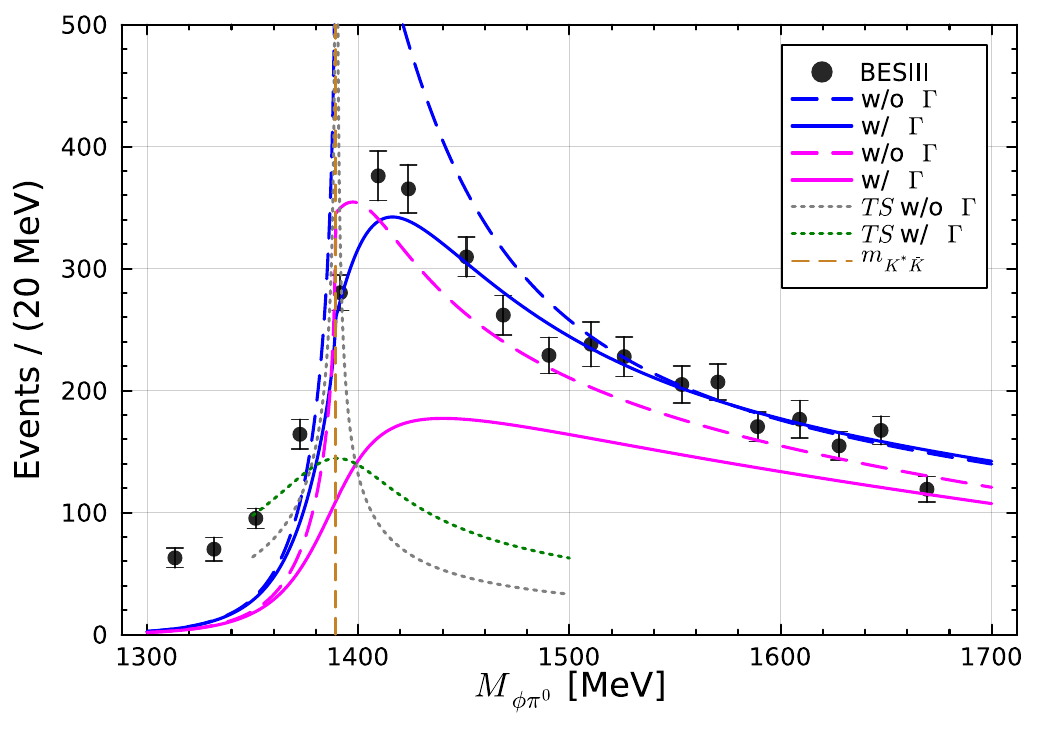}
	\caption{\maojun{Numerical results for Eq.~\eqref{Eq:dN_dpiphi} compared with the corresponding BESIII data~\cite{BESIII:2023zwx} for the $\phi\pi$ invariant-mass distribution. The blue and magenta curves have the same meaning as in Fig.~\ref{Amp_f0pi}, corresponding to $\alpha(\mu)=-1.63$ and $-1.28$, respectively. The dashed and solid curves denote the results without and with the finite $K^\ast$ width, respectively. The green and gray dotted curves show the TS-only contributions with and without the finite $K^\ast$ width, respectively. The vertical orange dashed line indicates the $K^\ast\bar K$ threshold.}
}\label{Amp_b1_Prediction}
\end{figure}

To test this hypothesis, we compare our model with the BESIII data. The theoretical expression is given by
\begin{equation}
\frac{d N_{b_1}}{d M_{\phi \pi}}=\frac{1}{32 \pi^3 M_{J / \psi}^2}\left|\mathcal{M}_{b_1}\right|^2\left|\vec{p}_\pi^*\right||\vec{k}|
\label{Eq:dN_dpiphi}
\end{equation}
with $N_{b_1}$ representing the number of events and $J/\psi$ mass $M_{J/\psi}$, 
where $|\vec{k}|$ denotes the magnitude of the c.m. momentum of the $\eta$ meson in the rest frame of the $J/\psi$, and $|\vec{p}_{\pi}^{\,*}|$ represents the momentum of the pion in the c.m. frame of the $\phi\pi$ pair. Furthermore, $\mathcal{M}_{b_1}$ represents the amplitude associated with the diagram in Fig.~\ref{b1_WT}, and given by
\begin{align}
\mathcal{M}_{b_1}=& P_{b_1} \hat{G}_{K^*\bar{K}}  T_{K^{\ast}\bar{K}\to K^{\ast}\bar{K}}\mathcal{M}^{TS}_{b_1}
\label{eq:b1_Product},
\end{align}
with 
\begin{eqnarray}
\mathcal{M}_{b_1}^{TS}&=&\int \frac{d^4 l}{\left( 2\pi\right)^4}\frac{\epsilon_i \delta^{ij}\left( 2p_{\pi}-l\right)_\maojun{j} \left(2l-P-p_{\pi}\right)^m\epsilon(\phi)_m}{\left( l^2-m_{K^{\ast}}^2\right) \left[ \left(P-l\right)^2-m_{\bar{K}}^2 \right] } \nonumber\\
&&\times \frac{1}{\left(l-p_{\pi} \right)^2-m_{K}^2 },\label{eq:b1TS}
\end{eqnarray}
where the effective Lagrangian in $\phi K\bar{K}$ vertex reads $\mathcal{L}_{\phi K\bar{K}}= g_1 \phi^{\mu}\left(K\partial_{\mu}\bar{K}-\partial_{\mu}K\bar{K}\right)$ and $P_{b_1}$ parameterizes the production of the $K^*\bar{K}$ pair, while $\hat{G}_{K^*\bar{K}}$, similar to the case in Section ~\ref{sec:IsovectorA}, is the $G$-loop function of the $K^*\bar{K}$ pair given by Eq.~\eqref{Eq:Gl_cut}. The $\epsilon_i$ and $\epsilon(\phi)$ are the polarizations of $b_1$ and $\phi$, respectively.
\maojun{The \(K^\ast\bar K\) scattering amplitude is treated using the on-shell factorization approximation, with the present analysis focusing on the near-threshold region. To maintain a consistent near-threshold treatment in the production amplitude, we retain only the leading contribution in the spatial loop momentum in the numerator of the triangle diagram, evaluated in the \(b_1\) rest frame. Specifically, we neglect the spatial loop momentum in the numerator while keeping its full momentum dependence in the propagator denominators. This approximation leads to a scalar triangle integral that is ultraviolet convergent and requires no additional regularization or subtraction. We emphasize that this is a near-threshold approximation rather than an exact consequence of the on-shell factorization of the scattering amplitude. Terms beyond this approximation may modify the energy dependence of the production amplitude, and the resulting line shape should therefore be interpreted qualitatively within the present framework.}


In Fig.~\ref{Amp_b1_Prediction}, we compare the line shape obtained from
Eq.~\eqref{Eq:dN_dpiphi} with the corresponding BESIII data
\cite{BESIII:2023zwx}. The dotted lines represent the TS-only
contributions and exhibit an enhancement around the
$K^{\ast}\bar K$ threshold. The green (gray) dotted line corresponds
to the result with (without) the finite $K^{\ast}$ width included in
the triangle diagram. As expected, the finite $K^{\ast}$ width
smears the narrow kinematic enhancement into a broader structure,
consistent with the behavior found in Ref.~\cite{Jing:2019cbw}.
Within the present setup, however, the TS-only contribution does not
reproduce the observed line shape as well as the result including
the $K^{\ast}\bar K$ final-state interaction.

When the nonperturbative $K^{\ast}\bar K$ rescattering is included, the results obtained with $\maojun{\alpha(\mu)=-1.28}$ (magenta) and $\maojun{\alpha(\mu)=-1.63}$ (blue) reproduce the main enhancement in the data around the $K^{\ast}\bar K$ threshold. The solid and dashed lines correspond to the results with and without the finite $K^\ast$ width, respectively. In particular, the blue solid curve captures the main enhancement around $1400\,\rm{MeV}$ and follows the overall trend of the data above the $K^{\ast}\bar K$ threshold. However, the agreement deteriorates below approximately $1350\,\rm{MeV}$, where contributions from the lower-lying $b_1(1235)$ and direct-production backgrounds may become important. Moreover, additional coupled channels may affect both the pole position and the line shape. These results suggest that the interplay between the TS kinematics and the pole-enhanced $K^{\ast}\bar K$ final-state interaction may play an important role in the observed enhancement, although the present single-channel framework does not provide a quantitative description of the entire $\phi\pi$ spectrum.

\maojun{For comparison with the BESIII data, the overall normalization of the full-amplitude curves is fixed using the data around \(M_{\phi\pi}\simeq 1700~\mathrm{MeV}\) as a reference, where the calculated amplitude varies relatively slowly with energy within the present model. The corresponding values of \(P_{b_1}\) are \(4.10\times10^{-3}~\mathrm{MeV}^{-1/2}\) for the blue solid and dashed curves and \(5.10\times10^{-3}~\mathrm{MeV}^{-1/2}\) for the magenta solid and dashed curves. For the TS-only curves, the normalization is chosen to make the calculated line shapes comparable to the BESIII data near the \(K^\ast\bar K\) threshold, with \(P_{b_1}=2.20\times10^{-3}~\mathrm{MeV}^{-1/2}\) for the gray dotted curve and \(3.10\times10^{-3}~\mathrm{MeV}^{-1/2}\) for the green dotted curve. These factors affect only the overall normalization and are chosen for illustrative comparison; no statistical fit to the energy dependence of the spectrum is performed.}

Below about $1350\,\rm{MeV}$, some deviations from the data remain.
This is not unexpected given the simplified nature of the present
model. First, the calculation employs the near-on-shell approximation
for the intermediate $K^{\ast}$, which becomes less reliable farther
below the $K^{\ast}\bar K$ threshold. Second, possible contributions
from the lower-lying $b_1(1235)$ are not included and may affect the
spectrum in this region. Direct tree-level background contributions
to the $\phi\pi$ final state are also neglected.

Despite these simplifications, the calculated line shape follows the
overall trend of the data and reproduces the enhancement around
$1400\,\rm{MeV}$. Together with the TS-only comparison above, this
suggests that the $K^{\ast}\bar K$ final-state interaction may play
an important role in shaping the observed near-threshold structure.

\maojun{Consequently, the present description is intended primarily for the near-\(K^\ast\bar K\)-threshold region, particularly the enhancement around \(1400~\mathrm{MeV}\), rather than for the entire \(\phi\pi\) invariant-mass spectrum. The possible contributions from coupled channels, the lower-lying \(b_1(1235)\), and direct production backgrounds may modify both the line shape and the pole position and should be investigated in a more comprehensive analysis.}


It is interesting to compare the second-sheet pole found here, which we
denote as $b_1(1400)$, with the $Z_s$ state proposed in
Ref.~\cite{Chen:2011cj} as a strange analogue of the $Z_c(3900)$.
The latter was motivated by the observation of the $Z_c(3900)$ by
BESIII in the $J/\psi\pi$ invariant-mass distribution of
$Y(4260)\to J/\psi\pi^+\pi^-$
\cite{BESIII:2013ris}. A search for such a $Z_s$ state was subsequently
performed by BESIII in the $\phi\pi$ invariant-mass distribution of
$e^+e^-\to\phi\pi\pi$, but no significant signal was observed
\cite{BESIII:2018rdg}. Likewise, the lattice-QCD study of
Ref.~\cite{Woss:2019hse} did not find evidence for an additional state
in the $\phi\pi$ channel that could be associated with the proposed
$Z_s$.

It should also be noted that the lattice calculation of
Ref.~\cite{Woss:2019hse} was performed at an unphysical pion mass,
$m_\pi\simeq391\,\rm{MeV}$. At this pion mass, the kaon and
$K^\ast$ masses are approximately $549\,\rm{MeV}$ and
$934\,\rm{MeV}$, respectively, so that the $K^{\ast}\bar K$
threshold is shifted substantially from its physical value.
Moreover, the pseudoscalar decay constant entering the WT interaction
is also modified. Since a near-threshold pole is generally sensitive
to both the interaction strength and the position of the corresponding
threshold, its location at this unphysical pion mass need not coincide
with that obtained here using physical hadron masses.

Furthermore, Ref.~\cite{Woss:2019hse} found that the relevant
axial-vector resonance couples predominantly to the $S$-wave
$\pi\omega$ channel, whereas its coupling to $\pi\phi$ is strongly
suppressed, consistent with the OZI rule. This suggests that the
$\phi\pi$ channel alone may have limited sensitivity to the
axial-vector dynamics in this energy region.

The pole discussed in the present work, however, is generated
predominantly by the near-threshold $K^{\ast}\bar K$ interaction.
A more direct lattice test would therefore require a dedicated
coupled-channel finite-volume analysis including the
$K^{\ast}\bar K$ channel and operators with good overlap onto this
configuration. Consequently, the absence of an additional signal
in Ref.~\cite{Woss:2019hse} does not by itself exclude the
near-threshold $K^{\ast}\bar K$ dynamics discussed here.

\section{EXTENSION TO OTHER FLAVOR SECTORS}
\label{sec:Isoscalar}

The near-threshold $K^{\ast}\bar K$ dynamics discussed above may also
have implications for other axial-vector states in the light-hadron
spectrum. In particular, attractive vector-pseudoscalar interactions
are relevant to the isoscalar $f_1$ and $h_1$ sectors, where the
$K^{\ast}\bar K$ threshold lies close to experimentally observed
structures. In this section, we briefly discuss these possible
connections, as well as the related $K^{\ast}K$ interaction in the
strange sector. These considerations are intended as qualitative
extensions of the physical picture developed in the isovector sector,
rather than as quantitative coupled-channel analyses.

\subsection{$f_1(1420)$}

In the isoscalar positive-$G$-parity sector, the WT interaction in the
$K^{\ast}\bar K$ channel is strongly attractive, with $C=-3$.
Consequently, the $K^{\ast}\bar K$ dynamics may play an important role
in the structure of the $f_1(1420)$. More generally, the physical
$f_1$ states need not correspond to either pure $q\bar q$ states or
pure hadronic molecules. Mixing between the bare flavor-singlet and
flavor-octet members of the $1^3P_1$ nonet and meson-meson components
provides a natural framework for describing such states
\cite{Ortega:2016mms,Albaladejo:2018mhb,Yan:2018zdt,Yang:2021tvc,
Peng:2023lfw,Wang:2023ovj,Shen:2025qpj,Wang:2025jcq,
Dai:2023kwv,Song:2023pdq}.

The relevance of the $K^{\ast}\bar K$ channel is also suggested by
the pronounced structure around $1.4\,\rm{GeV}$ observed in
$K\bar K\pi$ invariant-mass distributions, including those reported
by the OBELIX and DELPHI collaborations
\cite{OBELIX:2002eai,DELPHI:2003bnm}. Previous studies have emphasized
the strong coupling of the $f_1(1420)$ to the $K^{\ast}\bar K$
channel and the associated near-threshold dynamics
\cite{Debastiani:2016xgg,Liang:2017ijf,Hu:2026gbf}.
These observations suggest that the $f_1(1420)$ provides another
example in which a nearby hadronic threshold and intrinsic
$q\bar q$ components may both be relevant to the observed spectrum.

\subsection{$h_1(1415)$}

A similar situation occurs in the isoscalar negative-$G$-parity
sector. The attractive interaction associated with the
$K^{\ast}\bar K$ channel can generate pronounced near-threshold
dynamics and is therefore potentially relevant to the $h_1(1415)$,
formerly referred to as the $h_1(1380)$, which has been observed in
several experiments
\cite{Aston:1987ak,CrystalBarrel:1997kda,ParticleDataGroup:2022pth}.
As in the $f_1$ sector, the physical state may involve an interplay
between the conventional $^1P_1$ $q\bar q$ configuration and
meson-meson continuum components \cite{Clymton:2024pql}.
A quantitative determination of their relative importance would,
however, require a dedicated coupled-channel analysis and is beyond
the scope of the present work.

\subsection{Isoscalar $K^{\ast}K$ scattering}

It is also instructive to consider the related $K^{\ast}K$
interaction. A sufficiently attractive interaction in the isoscalar
channel could, in principle, generate a near-threshold state with
$\bar s\bar s ud$ flavor content, conventionally denoted as $T_{ss}$.
Such a state has been discussed as a light-flavor analogue of the
$T_{cc}$, whose near-threshold dynamics is closely related to the
$DD^{\ast}$ interaction
\cite{Meng:2021jnw,Ling:2021bir,Feijoo:2021ppq,Fleming:2021wmk,
Yan:2021wdl,Ren:2021dsi,Du:2021zzh,Albaladejo:2021vln,
Deng:2021gnb}.
For $I=0$, the corresponding $K^{\ast}K$ configuration can be written
as
\begin{align}
\vert \bar{s}\bar{s}ud\rangle
=
\frac{1}{\sqrt{2}}\vert K^{\ast +}K^0\rangle
-\frac{1}{\sqrt{2}}\vert K^{\ast 0}K^+\rangle .
\end{align}

Interestingly, the WT interaction for this $K^{\ast}K$ configuration
vanishes. Therefore, in contrast to the $K^{\ast}\bar K$ systems
discussed above, the leading WT interaction alone cannot generate a
$T_{ss}$ pole, and additional dynamical mechanisms would be required.
One possible contribution is axial-vector-meson exchange. Motivated
by the role of $a_1(1260)$ exchange discussed in studies of the
$Z_c$ spectrum \cite{Yan:2021tcp,Chen:2024bre,Chen:2026zdj},
we estimate the contribution from $a_1$ exchange to the
$K^{\ast}K$ interaction.

Using the coupling of the near-threshold $a_1$ pole to the
$K^{\ast}\bar K$ channel obtained in Sec.~\ref{sec:Isovector}, the
corresponding interaction is parametrically estimated as
\begin{align}
V^{a_1}
\simeq
\mathcal{O}\left(\frac{g_{a_1}^2}{m_{a_1}^2}\right),
\end{align}
where $g_{a_1}$ is of the order of $5\,\rm{GeV}$. At threshold, its
strength is estimated to be
\begin{align}
\left|V^{a_1}\right|
\sim
\frac{1}{10}
\left|V_{K^{\ast}\bar K}^{I=1}\right|.
\end{align}
This contribution is therefore substantially weaker than the
attractive $K^{\ast}\bar K$ interaction considered in the previous
sections and is also weaker than the $K^{\ast}K$ interactions obtained
in quark-model and one-boson-exchange studies
\cite{Ji:2024znj,Liu:2025fpx,Wang:2024kke,Tang:2025bcc}.
Within this estimate, $a_1$ exchange alone does not appear sufficient
to generate a near-threshold $T_{ss}$ pole. Whether additional
short-range or meson-exchange interactions can produce such a state
remains an open question deserving further theoretical and
experimental investigation.

\section{Summary}\label{sec:summary}

Using the chiral unitary approach, we have investigated the
near-threshold $K^{\ast}\bar K$ interaction and its implications for
the isovector axial-vector spectrum. The leading-order
WT interaction is unitarized through the
Bethe--Salpeter equation, generating second-sheet poles below the
$K^{\ast}\bar K$ threshold in both the positive- and negative-$G$-parity
sectors within the present single-channel framework.

For $G=\maojun{-}1$, we find a second-sheet pole below the
$K^{\ast}\bar K$ threshold associated with strong
$K^{\ast}\bar K$ dynamics. We have studied its possible manifestation
in the $f_0(980)\pi$ invariant-mass distribution measured by COMPASS,
where the $a_1(1420)$ structure was observed
\cite{COMPASS:2018uzl}. The TS generated by the
$K^{\ast}\bar K K$ loop produces an enhancement
near the $K^{\ast}\bar K$ threshold, while the nonperturbative
$K^{\ast}\bar K$ final-state interaction substantially modifies
the resulting line shape. Within the present framework, the
$a_1(1420)$ enhancement can be qualitatively understood through
the interplay between the TS mechanism and the pole-enhanced
$K^{\ast}\bar K$ final-state interaction
\cite{Mikhasenko:2015oxp,Aceti:2016yeb}. A quantitative description
of the full COMPASS spectrum would require a more complete treatment
of the relevant coupled channels and production dynamics.

In the $G=\maojun{+}1$ sector, the same underlying interaction
generates a partner second-sheet pole that may be associated with
the $b_1(1400)$ structure. We have examined its possible manifestation
in the $\phi\pi$ invariant-mass distribution measured by BESIII in
$J/\psi\to\eta\phi\pi$ \cite{BESIII:2023zwx}. As in the
$a_1(1420)$ case, the TS mechanism alone produces a near-threshold
enhancement, whereas the inclusion of the nonperturbative
$K^{\ast}\bar K$ rescattering qualitatively reproduces the main
enhancement around $1400\,\rm{MeV}$. However, the agreement with
the data deteriorates below approximately $1350\,\rm{MeV}$, where
contributions from the lower-lying $b_1(1235)$ and direct-production
backgrounds may be important. Furthermore, additional coupled
channels may modify both the pole position and the line shape.
The correlated behavior in the two $G$-parity sectors therefore
suggests that strong $K^{\ast}\bar K$ dynamics may play an important
role in the axial-vector spectrum near threshold, although the
present results do not establish a definitive identification of
the two structures as molecular states.

We have also briefly discussed the extension of this picture to other
flavor sectors. The strong isoscalar $K^{\ast}\bar K$ interaction is
potentially relevant to the $f_1(1420)$ and $h_1(1415)$, where
meson-meson dynamics may coexist with conventional $q\bar q$
components. In contrast, the leading WT interaction vanishes in the
isoscalar $K^{\ast}K$ channel, and our estimate indicates that
$a_1$ exchange alone is insufficient to generate a near-threshold
$T_{ss}$ pole. These observations motivate further experimental and
theoretical studies of axial-vector and related near-threshold
structures in the light-hadron spectrum.

\begin{acknowledgments}
We would like to thank Prof. Feng-Kun Guo for his suggestion on this study and Dr. J. M. Dias for his contributions at the early stage of this project. We also thank Profs. E. Oset, Wei-Hong Liang, and Chu-Wen Xiao for their valuable discussions.
This research is supported by the National Natural Science Foundation of China under Grants No. 12305096, the Fundamental Research Funds for the Central Universities under Grant
No. SWU-XDJH202304, No. SWU-KQ25016 and Chongqing Natural Science Foundation under Project No. CSTB2025NSCQ-GPX0516.
\end{acknowledgments}




\begin{thebibliography}{93}
\bibitem [{Zha()}]{Zhao:2013}%
  \BibitemOpen
  \href@noop {} {}\bibinfo {note} {The possible explanation of the $a_1(1420)$
  as a TS effect was first pointed out by Qiang Zhao at the HADRON2013
  Conference where the COMPASS result was first reported.}\BibitemShut {Stop}%
\end{thebibliography}

%

\end{document}